\documentclass{article}
\usepackage[all]{xy}
\usepackage{amssymb, amsmath}
\usepackage{accents}
\usepackage{cite}
\usepackage{subfigure}

\usepackage{graphicx}
\usepackage[caption=false]{subfig}

\usepackage{graphicx}
\usepackage[caption=false]{subfig}

\begin{document}

%\begin{frontmatter}

\title{Exact solution of the $p+ip$ Hamiltonian revisited: \\ duality relations in the hole-pair picture}
\author{Jon Links, Ian Marquette, Amir Moghaddam \\
School of Mathematics and Physics, The University of Queensland, \\ Brisbane, QLD 4072, Australia \\
email: jrl@maths.uq.edu.au, i.marquette@uq.edu.au, a.moghaddam1@uq.edu.au }

\date{}

\maketitle

\begin{abstract}
We study the exact Bethe Ansatz solution of the $p+ip$ Hamiltonian in a form whereby quantum numbers of states refer to hole-pairs, rather than particle-pairs used in previous studies. We find an asymmetry between these approaches. For the attractive system states in the strong pairing regime take the form of a quasi-condensate involving two distinct hole-pair creation operators. An analogous feature is not observed in the particle-pair picture.     
\end{abstract}
%\begin{keyword}
%Integrable systems \sep Bethe Ansatz \sep pairing model   
%\end{keyword}

%\end{frontmatter}

\section{Introduction}

The $p_x+ip_y$-wave pairing (or simply $p+ip$) Hamiltonian rose to prominence through the influential work of Read and Green \cite{rg00} in identifying topological properties of superconducting systems. An exact Bethe Ansatz solution of the Hamiltonian appeared in \cite{ilsz09}, a result which subsequently generated several studies in the exactly solvable framework \cite{s09,dilsz10,rdo10,onds13,ris14,vdv14,cdvv15}. Unlike its counterpart the exactly solvable $s$-wave pairing Hamiltonian, also known as the Richardson model \cite{r63,vdr01}, the $p+ip$ model  exhibits quantum phase transitions. A key to gaining a complete understanding of the model's properties is to understand the distribution of the roots of the Bethe Ansatz equations. For example, it is well-established \cite{dilsz10,rdo10} that there is a duality between states in the regimes known as {\it weak pairing} and {\it strong pairing}. There are states in the weak pairing regime characterised by the presence of particle-pairs which carry zero energy, and corresponding dual states in the strong pairing regime with the same energy. Particular groupings of particle-pairs should be viewed as bound, and only arise when certain constraints are satisfied. The existence of these zero energy particle-pairs may be interpreted as a form of quasi-condensation. 

Our objective is to revisit the Bethe Ansatz solution of the $p+ip$ model from the hole-pair perspective rather than the particle-pair perspective. This involves working with a very closely related, yet distinct, second form of Bethe Ansatz solution. There are a couple of motivations for taking this approach. The first is that Hirsch has long advocated that ``electron-hole asymmetry is the key to superconductivity'' \cite{h03}. As will be established, there is indeed an asymmetry between hole-pairs and particle-pairs which manifests in the character of the roots of the associated Bethe Ansatz equations. The second motivation is that there have been some very interesting studies of the Bethe Ansatz solutions for pairing Hamiltonians which exploit the existence of two forms of solutions. Pogosov and collaborators have determined sets of relations between the roots of the Bethe Ansatz equations in the $s$-wave \cite{plm13} and ``Russian Doll'' \cite{pb15} models, which in particular has led to a formula for the ground-state energy in the $s$-wave case. Also Faribault and collaborators \cite{fs12,tf14} have used the existence of the two forms of exact solution to facilitate the calculation of wavefunction overlaps and scalar products for a general class of systems, which includes the $s$-wave model.   Subsequently Claeys et al. \cite{cdvv15} have generalised this latter work to accommodate the $p+ip$ model.           

When the hole-pairing picture is adopted, we find that zero energy hole-pairs arise which characterise the same duality that was mentioned above. We also find that in addition there exist states characterised by {\it infinite} energy hole-pairs, in a manner that the sum of the infinite energies is finite. Such infinite energy solutions have been  previously observed \cite{onds13,vdv14}, but so far a systematic investigation of them has not been undertaken. Our main finding indicates a second form of duality, which relates the energies of eigenstates of the attractive pairing Hamiltonian to energies of eigenstates of the repulsive pairing Hamiltonian. It also provides a new perspective on the different regimes of the system in terms of hole-pair quasi-condensation, which we will discuss later.

In Sect. 2 we present the Hamiltonian and the known Bethe Ansatz solution in terms of particle-pair quantum numbers. We then determine the second solution in terms of hole-pair quantum numbers. Sect. 3 is devoted to the discussion of dualities. We commence by recalling the duality previously discussed in \cite{dilsz10,rdo10}. We then continue to establish a second form of duality, and ultimately a third which is a combination of the two. Sect. 4 examines these dualities in the framework of the Bethe Ansatz equations. In Sect. 5 we undertake numerical solution of the Bethe Ansatz equations, which leads into an investigation of the phase diagram in Sect. 6. Sect. 7 briefly compares the results of mean-field approximations with the results of the exact solution. Concluding remarks are given in Sect. 8.

\section{The Hamiltonian and exact solution}

We first introduce the Hamiltonian of the pairing model. We take the canonical (i.e. particle 
number-preserving) Hamiltonian whose mean-field approximation leads to the Bogoliubov-de Gennes equations with order  parameter having $p_x+ip_y$-wave symmetry up to quadratic approximation. Letting $c_{\bf k}$, $c_{\bf  k}^{\dagger}$ denote annihilation and creation operators for two-dimensional fermions of mass $m$ with momentum ${\bf k}=(k_{x},k_{y})$,   the Hamiltonian reads \cite{ilsz09}
\begin{align}
\mathcal{H}&=\sum_{{\bf k}}\frac{|\bf{k}|^{2}}{2m}c^\dagger_{{\bf k}}c_{\bf{k}} 
-\frac{G}{4m}\sum_{{\bf k}\neq \pm{\bf k}'}(k_{x}+ik_{y})(k'_{x}-ik'_{y})c_{{\bf k}}^{\dagger}
c_{-{\bf k}}^{\dagger}c_{-{\bf k}'}c_{{\bf k}'}, \nonumber
\end{align}
where $G$ is a dimensionless coupling constant which is positive for an attractive interaction and negative for a repulsive interaction.  
For any unpaired fermions the action of the pairing interaction is zero and we can thus decouple the Hilbert space into a product of paired and unpaired fermions states, for which  the action of the Hamiltonian on the space for the unpaired fermions is diagonal in the number operator basis. This is known as the {\it blocking effect} \cite{vdr01} and permits an analysis of a simplified version of the Hamiltonian.

We set $z_{{\bf k}}=|\textbf{k}|$ and $k_{x}+ik_{y}=|{\bf k}|\exp(i\phi_{{\bf k}})$. It is convenient to introduce the following phase-dependent Cooper pair (or {\it hardcore boson}) operators  ${b_{{\bf k}}}^{\dagger}=\exp(i\phi_{\bf k})c_{{\bf k}}^{\dagger}c_{{-\bf k}}^{\dagger}={b_{-{\bf k}}}^{\dagger}$,  $\tilde{b_{\bf k}}=\exp(-i\phi_{{\bf k}})c_{-{\bf k}}c_{{\bf k}}={b_{-\bf k}}$, and set ${N_{{\bf k}}}={b_{{\bf k}}}^{\dagger}{b_{{\bf k}}}=N_{-{\bf k}}$. 
Using integers to enumerate the unblocked pairs of momentum states, and setting $m=1$,  
the Hamiltonian takes the form 
\begin{align*}
H(G)&=\sum_{k=1}^Lz_k^2N_k-G\sum_{l=1}^L\sum_{k\neq l}^L z_l z_k b_l^\dagger b_k.  
\end{align*}
The hardcore boson operators satisfy the following commutation relations 
\begin{align}
 [b_{j},b_{k}^{\dagger}]=\delta_{jk}(I-2N_{j}),\qquad \qquad [b_{j},b_{k}]=[b_{j}^{\dagger},b_{k}^{\dagger}]=0  \label{comms}
\end{align}
where $I$ denotes the identity operator, as well as the relations $(b_{j}^{\dagger})^{2}=0$, $N_j^2=N_j$. The Hamiltonian may be expressed in the compact form 
\begin{align}
H(G)=(1+G)H_0-GQ^\dagger Q
\label{ham}
\end{align}
where 
\begin{align}
H_0&= \sum_{l=1}^Lz_l^2N_l,  &
Q^\dagger&=  \sum_{l=1}^L z_l b_l^\dagger,  &
Q&=  \sum_{l=1}^L z_l b_l.
\label{defs}
\end{align}
For later use we note the commutation relation 
\begin{align}
\left[Q^\dagger,\,Q\right]&=2H_0-\sum_{j=1}^Lz_j^2 I
\label{i1}
\end{align}
which follows from (\ref{comms}).

The Hamiltonian (\ref{ham}) is our principal object of study.
For each solution of the coupled equations
 \begin{align}
\frac{G^{-1}+2M-L-1}{\tilde{y}_k} + \sum_{l=1}^L\frac{1}{\tilde{y}_k-z_l^2}
& =\sum^M_{j\neq k}\frac{2}{\tilde{y}_k-\tilde{y}_j},
\qquad k=1,..., M  
\label{bae1}
\end{align}
there is an eigenstate of (\ref{ham}) with energy eigenvalue given by 
\begin{align} E=(1+G)\sum_{k=1}^M \tilde{y}_k. 
\label{nrg0}
\end{align}
The eigenstate has the form 
\begin{align*}
|\Phi\rangle = \prod_{j=1}^M C(\tilde{y}_j)|0\rangle
\end{align*}
where 
\begin{align*}
C(y)= \sum_{j=1}^L\frac{z_j}{y-z_j^2}b_j^\dagger.
\end{align*}
Above, $M$ is a quantum number which denotes the number of particle-pairs (i.e. Cooper pairs) for the state $|\Phi\rangle$, i.e.
\begin{align}
N|\Phi\rangle=M |\Phi\rangle.
\label{number}
\end{align} 

The exact solution was reported in \cite{ilsz09}, and subsequently shown that it could be derived through a variety of means including use of the classical Yang--Baxter equation \cite{s09}, the Quantum Inverse Scattering Method\cite{dilsz10}, or the Gaudin algebra \cite{rdo10}. A second form of exact solution can be obtained by a particle-hole transformation, denoted $\Upsilon$, which can be defined by 
\begin{align*}
\Upsilon N_l \Upsilon&=I-N_l, \\
\Upsilon b_l \Upsilon &=b_l^\dagger, \\
\Upsilon b_l^\dagger \Upsilon  &=b_l,
\end{align*}
where $\Upsilon=\Upsilon^{-1}$ and $I$ denotes the identity operator. The action of $\Upsilon$ naturally extends to states.  
In particular if $|\chi\rangle$ denotes the completely filled state of $L$ particle-pairs then 
\begin{align*}
|\chi\rangle=\Upsilon|0\rangle
\end{align*}
and if (\ref{number}) holds true then
\begin{align*}
N\Upsilon|\Phi\rangle=(L-M) \Upsilon|\Phi\rangle.
\end{align*} 
We then find that 
\begin{align}
\Upsilon H(G)\Upsilon =-H(-G)+\sum_{j=1}^L z_j^2 I .
\label{transform}
\end{align}
%From the observation that $\Upsilon^2$ acts as the identity operator, we can draw a number of conclusions regarding the %spectrum of $H(G)$. 
This provides a simple relationship between the spectrum of the attractive model with $G>0$ and that of the repulsive model with $G<0$.

Below, we will investigate in more detail the consequences of (\ref{transform}) in terms of the eigenstates.
First note that it follows that a second exact solution exists whereby
for each solution of the coupled equations 
\begin{align}
\frac{-G^{-1}+2P-L-1}{{y}_k} + \sum_{l=1}^L\frac{1}{{y}_k-z_l^2}& =\sum^P_{j\neq k}\frac{2}{{y}_k-{y}_j},
\qquad k=1,..., P  
\label{bae2}
\end{align}
there is an eigenstate of (\ref{ham}) with energy eigenvalue given by 
\begin{align}
E=\sum_{l=1}^Lz_l^2+(G-1)\sum_{k=1}^P {y}_k, 
\label{nrg}
\end{align}
where $P=L-M$ is the quantum number which denotes the number of hole-pairs.
In this instance the eigenstate is of the form 
\begin{align*}
|\Psi\rangle=\prod_{j=1}^P B(y_k)|\chi\rangle
\end{align*}
where
\begin{align}
B(y)&= \Upsilon C(y)\Upsilon  \nonumber \\
&= \sum_{j=1}^L\frac{z_j}{y-z_j^2}b_j.
\label{by}
\end{align}
This result can also be obtained by a direct calculation following the methods of \cite{bil12}. The details are provided in Appendix A.

\section{Duality}

Previous studies \cite{dilsz10,rdo10} have identified an exact duality relation in the spectrum of the Hamiltonian. Here we will first recall the essential aspects of that result, before continuing to establish a second exact duality.

A direct commutator calculation leads to the result
\begin{align*}
\left[H(G),\,C(0)\right]&= (1+G)\left[H_0,\,C(0)\right]-GQ^\dagger\left[Q,\,C(0)\right] \\
%&= -(1+G)Q^\dagger-GQ^\dagger(2N-L) \\
&=Q^\dagger(GL-2GN-G-1).
\end{align*}
Using proof by induction we have more generally
\begin{align}
\left[H(G),\,(C(0))^J\right]
&=JQ^\dagger(C(0))^{J-1}(GL-2GN-GJ-1)
\label{dressing1}
\end{align}
At this point it is important to make the observation that $N$ is a conserved operator, which partitions the space of states according to it eigenvalues $M=0,1,2,....$. Consider a state $|\Psi'\rangle$ satisfying 
\begin{align*}
H(G)|\Psi'\rangle &= E(G) |\Psi'\rangle, \\
N|\Psi'\rangle &= M' |\Psi'\rangle .
\end{align*} 
Choosing  $G^{-1}=L-2M-J$ and setting 
$$|\Psi\rangle = (C(0))^J |\Psi'\rangle $$
it follows from (\ref{dressing1}) that 
\begin{align*}
H(G)|\Psi\rangle &= E(G) |\Psi\rangle, \\
N|\Psi\rangle &= (M'+J) |\Psi\rangle .
\end{align*}
We call $|\Psi\rangle$ and $|\Psi'\rangle$ dual states. Setting $M=M'+J$ dual states are characterised by the relation  
\begin{align}
M+M'=L-G^{-1}.
\label{duality}
\end{align}
We may view the state $|\Psi\rangle$ as a quasi-condensate containing $J$ particle-pairs, each of zero energy in accordance with the expression (\ref{nrg0}).

In the hole-pair picture the duality relation (\ref{duality}) is maintained, and can be verified in a similar fashion. In this instance it is convenient to use (\ref{i1}) to express (\ref{ham}) as 
$$H=G\sum_{k=1}^Lz_k^2 I +(1-G) H_0-GQ Q^\dagger.$$ 
It is then found that 
\begin{align*}
\left[H(G),\,B(0)\right]&= (1-G)\left[H_0,\,B(0)\right]-GQ\left[Q^\dagger,\,B(0)\right] \\
%&= (1-G)Q-GQ(L-2N) \\
&=Q(2GN-GL-G+1).
\end{align*}
Using proof by induction we have, more generally,
\begin{align}
\left[H(G),\,(B(0))^J\right]
&=JQ(B(0))^{J-1}(2GN-GL-GJ+1).
\label{dressing2}
\end{align}
Consider a state $|\Phi'\rangle$ satisfying 
\begin{align*}
H(G)|\Phi'\rangle &= E(G) |\Phi'\rangle, \\
N|\Phi'\rangle &= M' |\Phi'\rangle .
\end{align*} 
Choosing  $G^{-1}=L-2M+J$ and setting 
\begin{align}
|\Phi\rangle = (B(0))^J |\Phi'\rangle 
\label{map00}
\end{align}
it follows from (\ref{dressing2}) that 
\begin{align*}
H(G)|\Phi\rangle &= E(G) |\Phi\rangle, \\
N|\Phi\rangle &= (M'-J) |\Phi\rangle .
\end{align*}
Setting $M=M'-J$ the states $|\Phi\rangle$ and $|\Phi'\rangle$ are characterised by the same duality relation (\ref{duality}).
Here we may view the state $|\Phi\rangle$ as a quasi-condensate containing $J$ hole-pairs, each of zero energy in accordance with the expression (\ref{nrg}).

\subsection{A mixed duality}

Next we turn the discussion toward a different form of duality relating eigenstates of the Hamiltonian $H(G)$ with eigenstates of $H(-G)$. 
We first note the preliminary lemma
\begin{align}
[H_0,\,Q^K]=KQ^{K-1}[H_0,\,Q]
\label{i2}
\end{align}
which is proved by induction. The calculation is straightforward so we omit the details. Expanding out the commutators in (\ref{i2}), this expression can be rewritten as 
\begin{align}
H_0Q^K-Q^KH_0&=KQ^{K-1}H_0Q-KQ^KH_0 \nonumber  \\
H_0Q^K+(K-1)Q^KH_0&=KQ^{K-1}H_0Q.
\label{i3}
\end{align}
Next we proceed to the following identity, which is also proved by induction:
\begin{align}
[H_0,\,Q^K]-[Q^\dagger, \,Q^K]Q=-K(H_0Q^K+Q^KH_0)+K\sum_{j=1}^Lz_j^2Q^K
\label{i4}
\end{align}
The case  $K=1$ may be verified directly using (\ref{i1}):
\begin{align}
[H_0,\,Q]-[Q^\dagger, \,Q]Q=-(H_0Q+QH_0)+\sum_{j=1}^Lz_j^2Q.
\label{i5}
\end{align}
Assuming that (\ref{i4}) holds true for $K=k-1$, and using (\ref{i5}), we obtain 
\begin{align*}
[H_0,\,Q^k]-[Q^\dagger, \,Q^k]Q&=[H_0,\,Q^{k-1}]Q+Q^{k-1}[H_0,\,Q]\\
&\qquad-[Q^\dagger, \,Q^{k-1}]Q^2-Q^{k-1}[Q^\dagger,\,Q]Q\\
&=\left(-(k-1)(H_0Q^{k-1}+Q^{k-1}H_0)+(k-1)\sum_{j=1}^Lz_j^2Q^{k-1}\right)Q  \\
&\qquad +Q^{k-1}\left(-(H_0Q+QH_0)+\sum_{j=1}^Lz_j^2Q\right)  \\
&=(1-k)H_0Q^k-kQ^{k-1}H_0Q-Q^k H_0+k\sum_{j=1}^L z_j^2 Q^k.
\end{align*}
Appealing to (\ref{i3}) with $K=k$ then 
yields 
$$[H_0,\,Q^k]-[Q^\dagger, \,Q^k]Q=-k(H_0Q^k+Q^k H_0)+k\sum_{j=1}^L z_j^2 Q$$
which establishes that (\ref{i4}) holds true for $K=k$. The result follows by induction. 

Finally,
\begin{align*}
H(G)Q^K+Q^KH(-G)&=\left((1+G)H_0-GQ^\dagger Q\right)Q^K+Q^K\left((1-G)H_0+GQ^\dagger Q\right)  \\
&=H_0Q^K+Q^KH_0 +G[H_0,\,Q^K]-G[Q^\dagger Q,\,Q^K] \\
&=H_0Q^K+Q^KH_0 +G[H_0,\,Q^K]-G[Q^\dagger,\,Q^K]Q.
\end{align*}
Through use of (\ref{i4}), this leads to the following identity:
\begin{align}
H(G)Q^K+Q^KH(-G)&=(1-KG)(H_0Q^K+Q^KH_0)+KG\sum_{j=1}^L z_j^2 Q^K.
\label{i0}
\end{align}

Consider a state $|\Theta'\rangle$ satisfying 
\begin{align*}
H(-G)|\Theta'\rangle &= E' |\Theta'\rangle, \\
N|\Theta'\rangle &= M' |\Theta'\rangle .
\end{align*} 
Choosing  $G^{-1}=K$ and setting 
\begin{align}
|\Theta\rangle = Q^K |\Theta'\rangle \label{map0}
\end{align}
it follows from (\ref{i0}) that 
\begin{align*}
H(G)|\Theta\rangle &= E |\Theta\rangle, \\
N|\Theta\rangle &= M |\Theta\rangle .
\end{align*}
where
\begin{align}
E&=\sum_{l=1}^L z_l^2-E', \label{nrgmixed} \\
M&=M'-K.  \nonumber
\end{align}
We call $|\Theta\rangle$ and $|\Theta'\rangle$ {\it mixed} dual states, which are  
characterised by  the relation 
\begin{align}
M'-M=G^{-1}.
\label{mixed}
\end{align}

Note that 
\begin{align*}
Q=\lim_{y\rightarrow \infty} y B(y).
\end{align*}
In contrast to the earlier discussed duality in terms of zero energy hole-pairs, the operator $Q$ is associated with an infinite energy hole-pair. However the energy of the state remains finite, as the sum of the diverging energies is finite. We prove this in Sect. \ref{unified} through an analysis of the Bethe Ansatz equations.

\subsection{The combined duality}
The two dualities described above can be combined to give the following result. Consider a state $|\Omega'\rangle$ satisfying
\begin{align*}
H(-G)|\Omega'\rangle&=E'|\Omega\rangle, \\
N|\Omega'\rangle&=M'|\Omega'\rangle.
\end{align*}
Setting
\begin{align}
|\Omega\rangle=Q^K (B(0))^J|\Omega' \rangle
\label{map}
\end{align}
where 
\begin{align*}
K&=G^{-1}, \\
J&=2M'-G^{-1}-L,  
\end{align*}
then
\begin{align*}
H(G)|\Omega\rangle&=E|\Omega\rangle, \\
N|\Omega\rangle&=M|\Omega\rangle,
\end{align*}
where
\begin{align*}
E&= \sum_{l=1}^L z_l^2-E', \\
M&=M'-K-J.
\end{align*}
The last relation is equivalent to 
\begin{align}
M+M'=L.
\label{combined}
\end{align}
 
One feature of the combined duality is that the dimension $d(M)$ of the sector for fixed $M$ is equal to $d(M')$. Explicitly, $d(M)$ is given by the binomial coefficient 
\begin{align}
d(M)&=\frac{L!}{M!(L-M)!}. 
\label{dim}
\end{align} 
It follows from (\ref{combined}) that $d(M)=d(M')$. This points to the possibility that the mapping from states $|\Omega'\rangle$ to $|\Omega\rangle$ through (\ref{map}) is a bijection, which we believe to be true. Numerical results which support this view are discussed in Sect. \ref{numerical}.

\section{Compatibility of dualities and Bethe Ansatz equations}
\label{unified}

In \cite{dilsz10} it was shown that the duality characterised by (\ref{duality})  is compatible with the Bethe Ansatz solution (\ref{bae1}). Here we extend that analysis to accommodate the second Bethe Ansatz solution (\ref{bae2}). 

Consider a generic splitting of the set of roots $Y$ of (\ref{bae2}), with $|Y|=P,$ into non-intersecting sets $Y'$ and $Z$ such that $Y=Y' \cup Z$. We can express (\ref{bae2}) as      
\begin{align}
&-G^{-1}+2P-L-1 +\sum_{l=1}^L \frac{y_k}{y_k-z_l^2}= \sum_{y_j\in Y', y_j\neq y_k} \frac{2y_k}{y_k-y_j}+ \sum_{y_j\in Z} \frac{2y_k}{y_k-y_j},   \quad y_k \in Y'  \label{A}
\\
&-G^{-1}+2P-L-1 +\sum_{l=1}^L \frac{y_k}{y_k-z_l^2} = \sum_{y_j\in Z, y_j\neq y_k} \frac{2y_k}{y_k-y_j}+\sum_{y_j\in Y'} \frac{2y_k}{y_k-y_j},    
\quad y_k \in Z. \label{B}
\end{align}    
Setting $|Y'|=S$ and $|Z|=T$ we take the sum in (\ref{B}) over elements in $Z$  to give
\begin{align}
T(-G^{-1}+2P-L-1)+\sum_{y_k\in Z}\sum_{l=1}^L \frac{y_k}{y_k-z_l^2} &= \sum_{y_j,y_k\in Z,y_j\neq y_k} \frac{2y_k}{y_k-y_j}+\sum_{y_j\in Y' ,y_k\in Z} \frac{2y_k}{y_k-y_j}
\nonumber \\
&= T(T-1)+\sum_{y_j\in Y' ,y_k\in Z} \frac{2y_k}{y_k-y_j}.    
\label{C}
\end{align}

Suppose that at some limiting value of $G$ we have 
\begin{align*}
y_k &\neq 0 \qquad{\rm for\,all}\,\,\, y_k \in Y , \\
y_k &=0 \qquad{\rm for\,all}\,\,\, y_k \in Z. 
\end{align*}
Taking note that $S+T=P$ 
Eq. (\ref{C}) informs us that 
\begin{align}
T=-G^{-1}+2P-L,
\label{em1}
\end{align} while from (\ref{A}) we obtain
\begin{align}
-G^{-1}+2P-L-1+\sum_{k=1}^L \frac{y_m}{y_m-z_k^2} &= \sum_{y_j\in Y', y_j\neq y_m} \frac{2y_m}{y_m-y_j}+2T,  && \quad y_m \in Y', \nonumber \\
-G^{-1}+2(P-T)-L-1+\sum_{k=1}^L \frac{y_m}{y_m-z_k^2} &= \sum_{y_j\in Y', y_j\neq y_m} \frac{2y_m}{y_m-y_j},  && \quad y_m \in Y'. \label{new1}
 \end{align}
Eq. (\ref{new1}) is the set of Bethe Ansatz equations for $S=P-T$ roots of an attractive system. The above calculations indicate that given such a solution set $Y'$, we can augment it with $T$ additional roots which all have zero value to obtain the solution set $Y$, provided that $T$ is given by (\ref{em1}). Identifying  
\begin{align*}
M&=L-P,   \\
M'&=L-S,
\end{align*}
then (\ref{em1}) is equivalent to the duality relation (\ref{duality}).

Alternatively, suppose that at some limiting value of $G$ we have 
\begin{align*}
y_m &\neq \infty \qquad{\rm for\,all}\,\, y_m \in Y' , \\
y_m &= \infty \qquad{\rm for\,all}\,\, y_m \in Z .
\end{align*}
Eq. (\ref{C}) in this instance leads to 
\begin{align} 
T=-G^{-1}+2P-2S,
\label{em2}
\end{align} 
while from (\ref{A}) we obtain
\begin{align}
-G^{-1}+2P-L-1 +\sum_{k=1}^L \frac{y_m}{y_m-z_k^2} &= \sum_{y_j\in Y', y_j\neq y_m} \frac{2y_m}{y_m-y_j},   \quad y_m \in Y' , \nonumber \\
G^{-1}+2(P-G^{-1})-L-1 +\sum_{k=1}^L \frac{y_m}{y_m-z_k^2} &= \sum_{y_j\in Y', y_j\neq y_m} \frac{2y_m}{y_m-y_j},   \quad y_m \in Y' . \label{new2}
 \end{align}
Eq. (\ref{new2}) is the set of Bethe Ansatz equations for $S=P-G^{-1}$ roots of a repulsive system. The above calculations indicate that given such a solution set $Y'$, we can augment it with $T$ additional roots which all have infinite value to obtain the solution set $Y$, provided that $T$ is given by (\ref{em2}), which simplifies to 
$T=G^{-1}$. Identifying as before
\begin{align*}
M&=L-P,   \\
M'&=L-S,
\end{align*}
then (\ref{em2}) is equivalent to the mixed duality relation (\ref{mixed}).

Moreover, from (\ref{B}) we find
\begin{align}
(-G^{-1}+2P-L-1) y_k +\sum_{l=1}^L \frac{y^2_k}{y_k-z_l^2} &= \sum_{y_j\in Z, y_j\neq y_k} \frac{2y^2_k}{y_k-y_j}+\sum_{y_j\in Y'} \frac{2y^2_k}{y_k-y_j},    
\quad y_k \in Z \nonumber \\
(-G^{-1}+2P-L-1) \sum_{y_k\in Z} y_k +\sum_{y_k\in Z}\sum_{l=1}^L \frac{y_k(y_k-z_l^2+z_l^2)}{y_k-z_l^2} &= \sum_{y_k,y_j\in Z, y_j\neq y_k} \frac{2y^2_k}{y_k-y_j}+\sum_{y_k\in Z,y_j\in Y'} \frac{2y_k(y_k-y_j+y_j)}{y_k-y_j} 
\nonumber \\
(-G^{-1}+2P-1) \sum_{y_k\in Z} y_k +\sum_{y_k\in Z}\sum_{l=1}^L \frac{y_k z_l^2}{y_k-z_l^2} &= 2(S+T-1)\sum_{y_k\in Z} y_k+\sum_{y_k\in Z,y_j\in Y'} \frac{2y_k y_j}{y_k-y_j}  \label{equ}
\end{align}
Considering the limit $G^{-1}\rightarrow T$, such that $y_k\rightarrow \infty$ for $y_k\in\,Z$, we let 
\begin{align*}
\sigma=\lim_{G^{-1}\rightarrow T} \sum_{y_k\in\,Z} y_k.
\end{align*}
Eq. (\ref{equ}) then reduces to
\begin{align*}
\sum_{k=1}^L z_k^2 &= (1-T^{-1})\sigma+2\sum_{y_j\in Y'} {y_j} 
\end{align*} 
and consequently the energy is found from (\ref{nrg}): 
\begin{align*}
E&= \sum_{l=1}^L z_l^2+\lim_{G^{-1}\rightarrow T}(G-1)\sum_{k=1}^P y_k \\
&=\sum_{l=1}^L z_l^2+(T^{-1}-1)\sum_{y_j\in Y'} {y_j} +(T^{-1}-1)\sigma \\
&= (1+T^{-1})\sum_{y_j\in Y'} {y_j}. 
\end{align*}
The above expression is simply (\ref{nrg0}), with the set $\{y_j\}$ a solution of (\ref{bae1}).
Proceeding further we have
\begin{align*}
E&=\sum_{l=1}^Lz_l^2-E'
\end{align*}
in agreement with (\ref{nrgmixed}), where 
$$E'=\sum_{l=1}^L z_l^2-(1+T^{-1})\sum_{y_j\in Y'} {y_j}$$
is the energy of the dual state for the repulsive Hamiltonian $H(-T^{-1})$ as given by (\ref{nrg}).

\section{Numerical results}
\label{numerical}

\begin{figure}[h]
		\centering
	\subfigure[$$]{
		\centering
		\includegraphics[width=.47\linewidth]{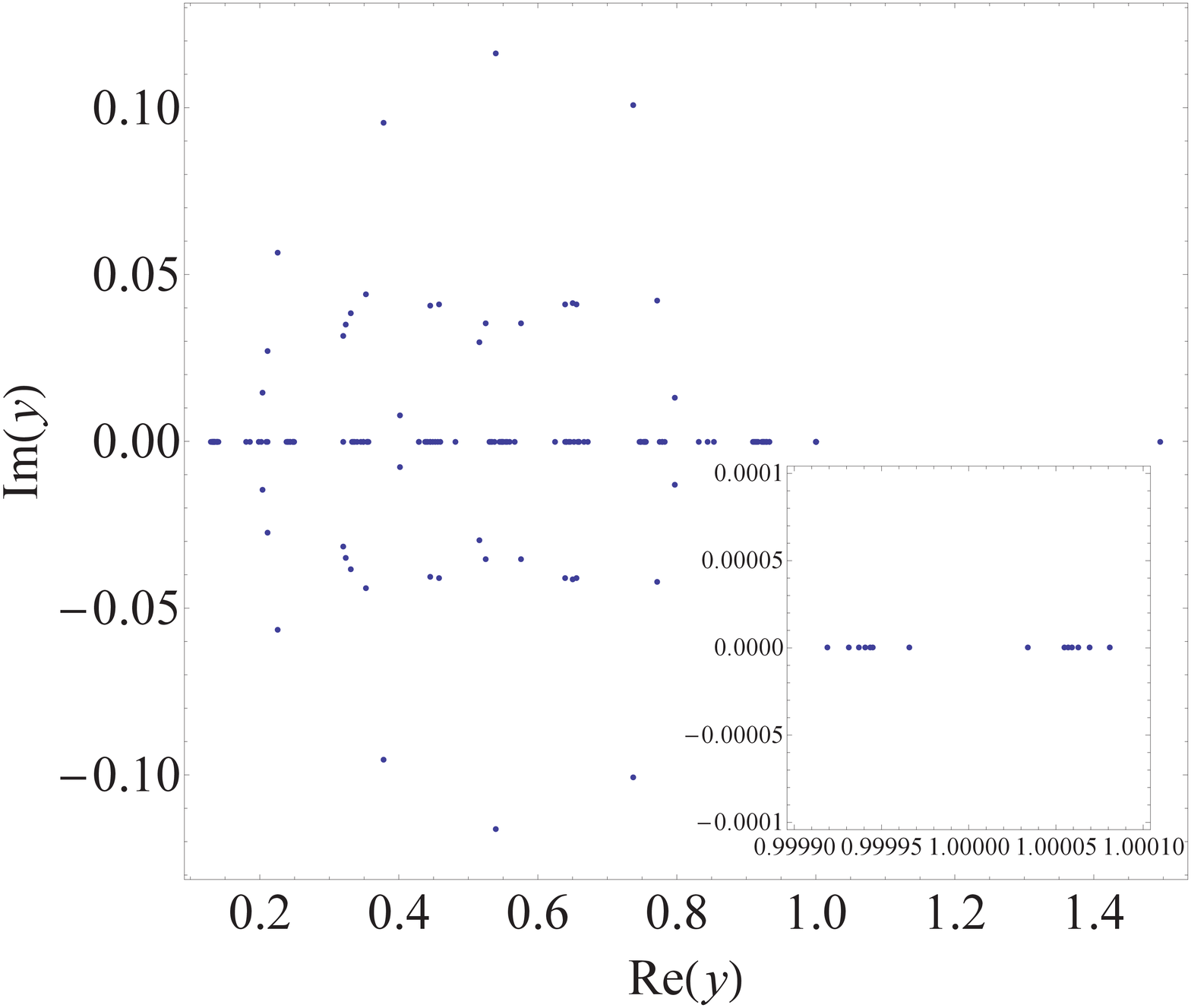}
}
	~
	\centering
	\subfigure[$$]{
		\centering
		\includegraphics[width=.47\linewidth]{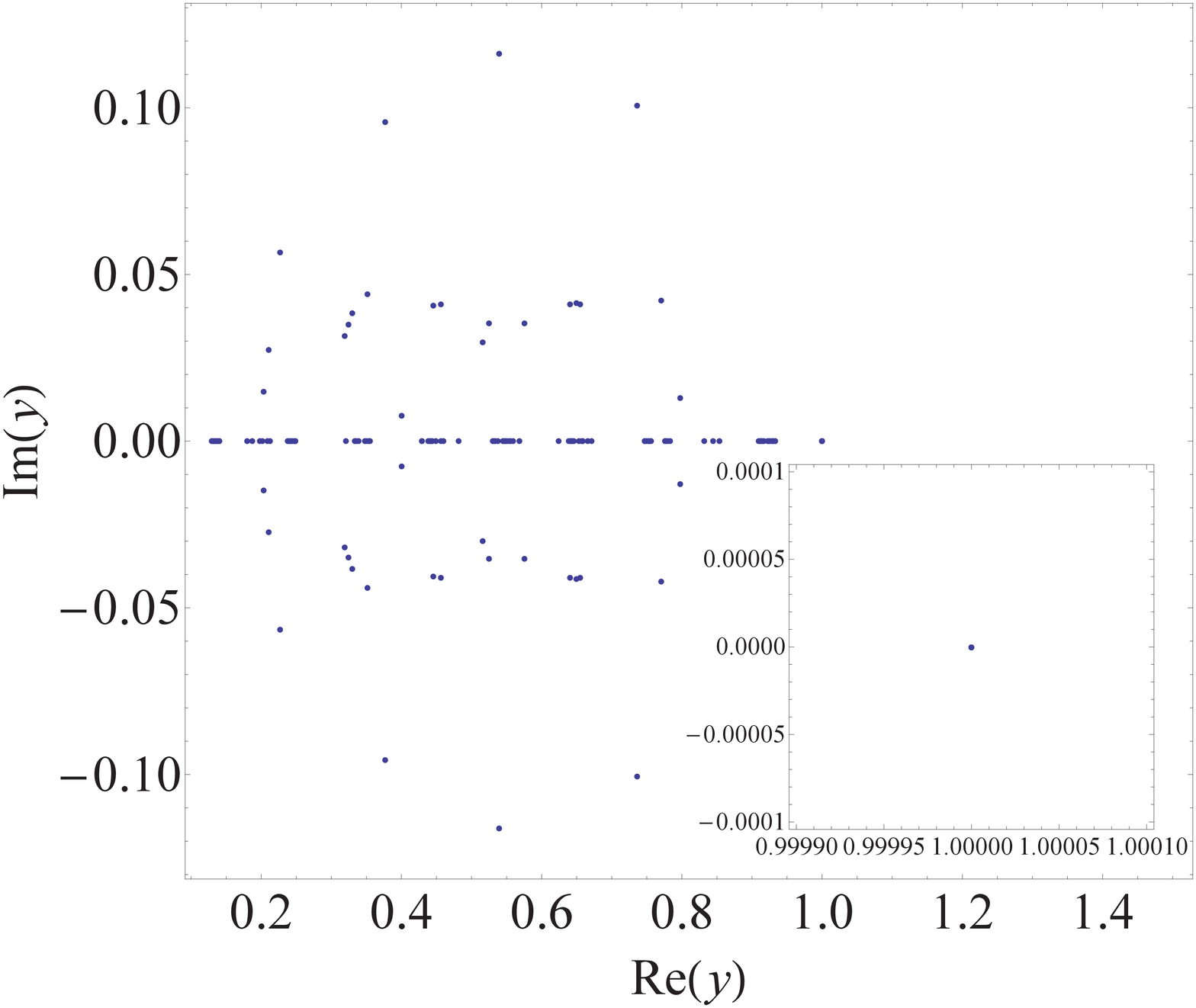}
%}\\
%	\subfigure[$M=3$]{
%		\centering
%		\includegraphics[width=.4\linewidth]{L=6_M=3}
%}
%	~
%	\subfigure[$M=4$]{
%		\centering
%		\includegraphics[width=.4\linewidth]{L=6_M=4}
%}\\
%	\subfigure[$M=5$]{
%		\centering
%		\includegraphics[width=.4\linewidth]{L=6_M=5}
}
 \caption{Distribution of roots for $L=8, P=3$, and $ \epsilon_j=j/10$. The panel (a) corresponds to the system with $G=1/2+0.0000001$. The inset in panel (a) shows the distribution of roots around Re$(y)=1$. As $G\rightarrow1/2$ sixteen roots  collapse to the point Re$(y)=1$, as illustrated in panel (b) and its inset for which $G=1/2$.}
\label{fig:L=6M=1-5}
\end{figure}
It has been established that the Bethe Ansatz solution (\ref{bae2}) has the property that for particular values of $G$ there are roots which are zero, and others which are infinite. To numerically solve (\ref{bae2}) near these values necessarily means that the elements of the solution set will vary across several orders of magnitude,  potentially imposing a large computational cost. To alleviate this issue we perform a change of variables:
\begin{align}
{y}_j&=\frac{1+v_j}{1-v_j}, \label{moebius1}\\
z_l^2&=\frac{1+\varepsilon_l}{1-\varepsilon_l}, \label{moebius2}
\end{align}  
such that
\begin{align*}
{y}_j&=0     & \Longleftrightarrow & &   v_j&= -1, \\
{y}_j&=\infty   &  \Longleftrightarrow  & & v_j&=1. 
\end{align*}
Under this change of variables the Bethe Ansatz equations (\ref{bae2}) become
\begin{align}
\frac{-2G^{-1}+4P-2L-2}{1-v^2_n} + \frac{L-2P+2}{1-v_n} +\sum_{l=1}^L\frac{1}{v_n-\varepsilon_l}& =
\sum^P_{q\neq n}\frac{2}{v_n-v_q} 
\label{transbae}
\end{align}
To numerically solve the above equations we adapt a technique described in \cite{ml12}.
 
The first case we consider is a system with $L=8$ and $P=3$. In this sector the dimension of the state space is 56. Consider $G=1/2$. With respect to Eq. (\ref{mixed}), the dual sector corresponds to $P=1$, which has dimension 8.  In this instance the mapping between sectors given by (\ref{map0}) cannot be a bijection.  
 
We perturb the coupling by a small amount and numerically solve (\ref{transbae}) for all roots, with the results displayed in Fig. 2.  It can seen that that there is a subset of roots close to the value 1. As $G\rightarrow 1/2$ they converge to 1.  In Table 1 the root sets are sorted according to increasing energy which is computed through (\ref{nrg}). It is apparent that the eight lowest energy states each have two roots close to the value 1. Between the particular sectors $P=3$ and $P=1$, the operator $Q^2$ is an injection in the limit $G\rightarrow 1/2$. However since the dimension of the co-domain is larger than the dimension of the domain, not all states in the co-domain are of the form (\ref{map}).
 
The second case we consider is a system with $L=8$ and $P=6$. In this sector the dimension of the state space is 28. Again consider $G=1/2$. With respect to Eq. (\ref{combined}), the dual sector corresponds to $P=2$ which also has dimension 28.  In this instance the mapping between sectors given by (\ref{map0}) may possibly be a bijection.  
 
We again perturb the coupling by a small amount to make it apparent that the solutions we obtain are not spurious, and that we can account for the full dimension of the sector. We numerically solve (\ref{transbae}), with the results  displayed in Table 2. The root sets are sorted according to increasing energy which is computed through (\ref{nrg}). It is apparent that {\it all} energy states each have a complex conjugate pair of roots close to the value $-1$, and two real roots close to the value 1. As $G\rightarrow 1/2$ they collapse to $-1$ and 1 respectively  (results not shown). Between these particular sectors $P=6$ and $P=2$ the operator $Q^2(B(0))^2$ is a bijection. All states in the co-domain are of the form (\ref{map}). Although we do not have a proof, we expect that this is true in general for sectors related through the combined duality (\ref{combined}). A necessary condition for this to be the case is that the mapping (\ref{map00}) is a surjection. This requires that $d(M')\geq d(M)$ where $d(M)$ denotes the dimension formula (\ref{dim}) and $M,\,M'$ are related through  (\ref{mixed}). A proof of this result is given in Appendix B.

\begin{table}\centering
\begin{tabular}{|c|l||c|l|}\hline
\multicolumn{4}{|c|}{$L=8, P=3, G =(1/2) + 0.0000001$, and $\varepsilon_j=j/10$} \\ [0.5ex]\hline\hline
Energy   & Bethe roots&Energy   & Bethe roots      \\ \hline\hline
$ -7.54299 $  &  $0.99997,1.00003,1.49644$ &  $ 22.36145 $  &  $0.35540,0.54943,0.74665$\\\hline
$ 1.94946 $  &  $0.13030,0.99994,1.00006$ &  $ 22.43743 $  &  $0.74865,0.45717\pm0.04099 \textit{i}$\\\hline
$ 2.44011 $  &  $0.23860,0.99994,1.00006$ &  $ 22.55229 $  &  $0.24377,0.55370,0.74773$\\\hline
$ 3.08380 $  &  $0.34552,0.99994,1.00006$ &  $ 22.70038 $  &  $0.13294,0.55556,0.74847$\\\hline
$ 3.97611 $  &  $0.45217,0.99994,1.00006$ &  $ 22.92889 $  &  $0.24797,0.44046,0.75208$\\\hline
$ 5.30314 $  &  $0.55903,0.99993,1.00007$ &  $ 23.05279 $  &  $0.35360,0.63967\pm0.04101 \textit{i}$\\\hline
$ 7.49744 $  &  $0.66657,0.99992,1.00008$ &  $ 23.07422 $  &  $0.75333,0.32407\pm0.03481 \textit{i}$\\\hline
$ 11.87227 $  &  $0.77566,0.99989,1.00011$ &  $ 23.07834 $  &  $0.13439,0.44448,0.75282$\\\hline
$ 12.52145 $  &  $0.18033,0.21200,0.93364$ &  $ 23.12185 $  &  $0.24304,0.65007\pm0.04141 \textit{i}$\\\hline
$ 12.64936 $  &  $0.13716,0.33338,0.93230$ &  $ 23.20321 $  &  $0.13263,0.65541\pm0.04094 \textit{i}$\\\hline
$ 12.83630 $  &  $0.93013,0.31915\pm0.03170 \textit{i}$ &  $ 23.33978 $  &  $0.13784,0.33456,0.75472$\\\hline
$ 12.87624 $  &  $0.13380,0.44202,0.92986$ &  $ 23.34662 $  &  $0.48232,0.53200,0.62467$\\\hline
$ 12.98062 $  &  $0.24692,0.43819,0.92852$ &  $ 23.49317 $  &  $0.18649,0.20946,0.75562$\\\hline
$ 13.24971 $  &  $0.13239,0.55018,0.92567$ &  $ 23.78106 $  &  $0.54736,0.54021\pm0.11610 \textit{i}$\\\hline
$ 13.31567 $  &  $0.92413,0.44485\pm0.04048 \textit{i}$ &  $ 24.17544 $  &  $0.24905,0.44448,0.63927$\\\hline
$ 13.34510 $  &  $0.24279,0.54830,0.92424$ &  $ 24.30757 $  &  $0.64194,0.33081\pm0.03821 \textit{i}$\\\hline
$ 13.47474 $  &  $0.35396,0.54410,0.92226$ &  $ 24.32987 $  &  $0.13496,0.44873,0.64022$\\\hline
$ 13.99730 $  &  $0.13162,0.65940,0.91663$ &  $ 24.60177 $  &  $0.13850,0.33606,0.64421$\\\hline
$ 14.07798 $  &  $0.24104,0.65843,0.91494$ &  $ 24.63370 $  &  $0.24842,0.51582\pm0.02981 \textit{i}$\\\hline
$ 14.16919 $  &  $0.91099,0.57513\pm0.03533 \textit{i}$ &  $ 24.73161 $  &  $0.13481,0.52553\pm0.03537 \textit{i}$\\\hline
$ 14.18711 $  &  $0.34973,0.65670,0.91260$ &  $ 24.75667 $  &  $0.19766,0.20219,0.64576$\\\hline
$ 14.34237 $  &  $0.45985,0.65262,0.90915$ &  $ 24.87852 $  &  $0.53021,0.35199\pm0.04392 \textit{i}$\\\hline
$ 17.62521 $  &  $0.13125,0.77844,0.85271$ &  $ 25.18880 $  &  $0.42933,0.37736\pm0.09555 \textit{i}$\\\hline
$ 17.77665 $  &  $0.24030,0.77968,0.84490$ &  $ 25.25408 $  &  $0.13948,0.33963,0.53465$\\\hline
$ 18.00125 $  &  $0.34831,0.78269,0.83171$ &  $ 25.40724 $  &  $0.53757,0.20335\pm0.01472 \textit{i}$\\\hline
$ 18.38397 $  &  $0.45676,0.79787\pm0.01289 \textit{i}$ &  $ 25.57900 $  &  $0.13973,0.40092\pm0.00763 \textit{i}$\\\hline
$ 19.32412 $  &  $0.56759,0.77091\pm0.04229 \textit{i}$ &  $ 25.79494 $  &  $0.42898,0.21141\pm0.02721 \textit{i}$\\\hline
$ 20.41664 $  &  $0.67105,0.73644\pm0.10065 \textit{i}$ &  $ 26.03673 $  &  $0.31998,0.22656\pm0.05659 \textit{i}$\\\hline
\end{tabular}
\caption {Numerical solution of the Bethe Ansatz Eqs. (\ref{transbae}) with $L=8, P=3, G =(1/2) + 0.0000001$ and $ \varepsilon_j=j/10$. The energies are calculated through (\ref{nrg}) after transforming back to the variables $y_j$ and $z_l^2$ through (\ref{moebius1},\ref{moebius2}).}
\end{table}

\newpage

\begin{table}\centering
\begin{tabular}{|c|l|}\hline
\multicolumn{2}{|c|}{$L=8, P=6, G =(1/2) + 0.0000001$ and $\varepsilon_j=j/10$} \\ [0.5ex]\hline\hline
Energy & Bethe roots \\ \hline\hline
$ -11.83981209 $  &  $ -0.99999981\pm 0.00024393 \textit{i}, 0.99997107, 1.00002897, 1.35407523, 9.17378076$\\ \hline
$ -3.65796172 $  &  $ -0.99999946\pm 0.00057037 \textit{i}, 0.99996003, 1.00004003, 1.72872448, 0.13265362$\\ \hline
$ -2.80008720 $  &  $ -0.99999953\pm 0.00050626 \textit{i}, 0.99995840, 1.00004166, 1.79893201, 0.24145319$\\ \hline
$ -1.74992090 $  &  $ -0.99999959\pm 0.00045582 \textit{i}, 0.99995624, 1.00004382, 1.89428154, 0.34849627$\\ \hline
$ -0.41519412 $  &  $ -0.99999963\pm 0.00041512 \textit{i}, 0.99995320, 1.00004686, 2.02752239, 0.45498685$\\ \hline
$ 1.37437992 $  &  $ -0.99999967\pm 0.00038118 \textit{i}, 0.99994857, 1.00005151, 2.21634859, 0.56145363$\\ \hline
$ 4.02553678 $  &  $ -0.99999970\pm 0.00035212 \textit{i}, 0.99994039, 1.00005969, 2.48406296, 0.66839900$\\ \hline
$ 4.55935661 $  &  $ -0.99999662\pm 0.00199582 \textit{i}, 0.99993382, 1.00006628, 0.20680832\pm 0.02043937 \textit{i}$\\ \hline
$ 5.01704811 $  &  $ -0.99999709\pm 0.00178555 \textit{i}, 0.99993146, 1.00006865, 0.14110366, 0.33689835$\\ \hline
$ 5.89937520 $  &  $ -0.99999774\pm 0.00148867 \textit{i}, 0.99992647, 1.00007365, 0.13730723, 0.44668753$\\ \hline
$ 6.01657120 $  &  $ -0.99999807\pm 0.00128240 \textit{i}, 0.99992550, 1.00007462, 0.33690194\pm 0.03941953 \textit{i}$\\ \hline
$ 6.38806868 $  &  $ -0.99999819\pm 0.00123967 \textit{i}, 0.99992318, 1.00007694, 0.25126848, 0.44251302$\\ \hline
$ 7.21930929 $  &  $ -0.99999827\pm 0.00124763 \textit{i}, 0.99991780, 1.00008233, 0.13570667, 0.55544129$\\ \hline
$ 7.70384297 $  &  $ -0.99999861\pm 0.00104596 \textit{i}, 0.99991363, 1.00008651, 0.24697377, 0.55356595$\\ \hline
$ 8.18565261 $  &  $ -0.99999896\pm 0.00084940 \textit{i}, 0.99990850, 1.00009165, 0.46637678\pm 0.03850363 \textit{i}$\\ \hline
$ 8.33050601 $  &  $ -0.99999890\pm 0.00088637 \textit{i}, 0.99990761, 1.00009255, 0.35820837, 0.54928501$\\ \hline
$ 8.82488231 $  &  $ -0.99999973\pm 0.00032685 \textit{i}, 0.99992267, 1.00007744, 2.86124262, 0.77673107$\\ \hline
$ 9.40566610 $  &  $ -0.99999862\pm 0.00107506 \textit{i}, 0.99989980, 1.00010037, 0.13481277, 0.66436225$\\ \hline
$ 9.88812763 $  &  $ -0.99999887\pm 0.00091253 \textit{i}, 0.99989340, 1.00010678, 0.24509789, 0.66345460$\\ \hline
$ 10.51623429 $  &  $ -0.99999908\pm 0.00078718 \textit{i}, 0.99988379, 1.00011642, 0.35409601, 0.66184154$\\ \hline
$ 11.36752125 $  &  $ -0.99999923\pm 0.00069725 \textit{i}, 0.99986777, 1.00013248, 0.46378746, 0.65803096$\\ \hline
$ 11.72101631 $  &  $ -0.99999930\pm 0.00065253 \textit{i}, 0.99985675, 1.00014354, 0.59339982\pm 0.01996575 \textit{i}$\\ \hline
$ 13.77136379 $  &  $ -0.99999885\pm 0.00095049 \textit{i}, 0.99985955, 1.00014067, 0.13423675, 0.77454140$\\ \hline
$ 14.25217295 $  &  $ -0.99999905\pm 0.00081659 \textit{i}, 0.99984872, 1.00015152, 0.24402831, 0.77416049$\\ \hline
$ 14.88038709 $  &  $ -0.99999921\pm 0.00071424 \textit{i}, 0.99983265, 1.00016762, 0.35222296, 0.77357060$\\ \hline
$ 15.74513536 $  &  $ -0.99999932\pm 0.00063925 \textit{i}, 0.99980693, 1.00019339, 0.46018824, 0.77251429$\\ \hline
$ 17.00197794 $  &  $ -0.99999940\pm 0.00058367 \textit{i}, 0.99976580, 1.00023458, 0.56880755, 0.77001849$\\ \hline
$ 18.42439918 $  &  $ -0.99999946\pm 0.00054548 \textit{i}, 0.99971075, 1.00028967, 0.68228619, 0.74962317$\\ \hline
\end{tabular}
\caption {Numerical solution of the Bethe Ansatz Eqs. (\ref{transbae}) with $L=8, P=6, G =(1/2) + 0.0000001$ and $ \varepsilon_j=j/10$. The energies are calculated through (\ref{nrg}) after transforming back to the variables $y_j$ and $z_l^2$ through (\ref{moebius1},\ref{moebius2}).}
\end{table}

\section{Phase diagram}

We introduce the rescaled coupling parameter $g=GL$ and the filling fraction $x=M/L$. It is convenient to represent the duality relations in terms of the phase diagram shown in Fig. 2, which depicts six regions in the $g^{-1}-x$ plane. For the attractive model with $g>0$ the three regions denoted IV, V, and VI have been previously identified as the strong pairing, weak pairing, and weak coupling regimes respectively \cite{dilsz10}. 
The weak pairing and strong pairing regimes are dual with respect to (\ref{duality}). The boundary between these regions is known as the Read-Green line and is given by the relation
\begin{align*}
x=\frac{1}{2}(1-g^{-1}).
\end{align*}
This line extends into the repulsive region $g<0$ and provides the boundary between regions II and III which are also dual with respect to (\ref{duality}). 
The boundary between the regions IV and VI is known as the Moore-Read line, and is given by 
\begin{align*}
x=1-g^{-1}.
\end{align*}
The ground state on the Moore-Read line is dual to the vacuum through (\ref{duality}) and has zero energy. 

With respect to the mixed duality relation (\ref{mixed}) regions II and V are dual, as are regions III and IV. With respect to the combined duality governed by (\ref{combined}), regions III and V are dual. 
For a state in the strong pairing regime (region V) which may  be expressed in the form (\ref{map}), we define the fraction of zero energy hole-pairs as 
\begin{align*}
h_0=\frac{J}{L}
\end{align*}
and the fraction of infinite energy hole-pairs as 
\begin{align*}
h_\infty=\frac{K}{L}
\end{align*}
where as before $P=L-M$.
The states (\ref{map}) only exist for certain integer values of $G^{-1}$, however in the thermodynamic limit the values of $h_0$ and $h_\infty$ become dense.   
The thermodynamic limit is obtained by 
taking the limits 
\begin{align*}
M&\rightarrow \infty, \\
L&\rightarrow \infty, \\
G&\rightarrow 0
\end{align*}
such that $x$ and $g$ are finite \cite{dilsz10,rdo10}.  In this limit we have 
\begin{align*}
h_0&= 1-2x-g^{-1}, \\
h_\infty &= g^{-1} 
\end{align*}
such that $h_0+h_\infty$ is independent of $g$. 

If we conduct an analogous analysis in the particle-pair picture, through use of the Bethe Ansatz solution (\ref{bae1}), we do not obtain a complementary portrait. There are no zero energy particle-pairs, nor are there any infinite energy particle-pairs, in the strong pairing regime. This is a key result of this study, that there is a clear asymmetry between the hole-pair picture and the particle-pair picture.

%Similarly, in the weak pairing regime (region IV) we have 
%\begin{align*}
%h_0&= 0, \\
%h_\infty &= g^{-1} .
%\end{align*}

\subsection{Inversion}
It is worth briefly mentioning that besides the duality relations discussed above, there exists another type of relation which we call inversion. Consider the Bethe Ansatz Eqs. (\ref{bae2}) and set $u_k=y_k^{-1}$. Then
 \begin{align*}
(-G^{-1}+2P-L-1)u_k + \sum_{l=1}^L\frac{u_kz_l^{-2}}{z_l^{-2}-u_k}& =\sum^P_{j\neq k}\frac{2u_ku_j}{{u}_j-{u}_k} \\
-G^{-1}+2P-L-1 + \sum_{l=1}^L\frac{z_l^{-2}-u_k+u_k}{z_l^{-2}-u_k}& =\sum^P_{j\neq k}\frac{2(u_j-u_k+u_k)}{{u}_j-{u}_k}  \\
-G^{-1}+2P-1 + \sum_{l=1}^L\frac{u_k}{z_l^{-2}-u_k}& =2P-2+\sum^P_{j\neq k}\frac{2u_k}{{u}_j-{u}_k} \\
\frac{G^{-1}-1}{u_k} + \sum_{l=1}^L\frac{1}{u_k-z_l^{-2}}& =\sum^P_{j\neq k}\frac{2}{{u}_k-{u}_j} \\
\frac{-\tilde{G}^{-1}+2P-L-1}{u_k} + \sum_{l=1}^L\frac{1}{u_k-z_l^{-2}}& =\sum^P_{j\neq k}\frac{2}{{u}_k-{u}_j}
\end{align*}
where 
$\tilde{G}^{-1}=-G^{-1}+2P-L$. If the momentum parameters are chosen such that $\{z_l:l=1,...,L\}=\{z_l^{-1}:l=1,...,L\}$ inversion maps roots for a Hamiltonian $H(G)$ to a set of roots for the Hamiltonian $H(\tilde{G})$. It provides a invertible mapping between solutions sets in regions I and VI, between solutions sets in regions II and IV, while regions III and V are each stable under inversion. 

\begin{figure}[h]
{\includegraphics[width=16cm]{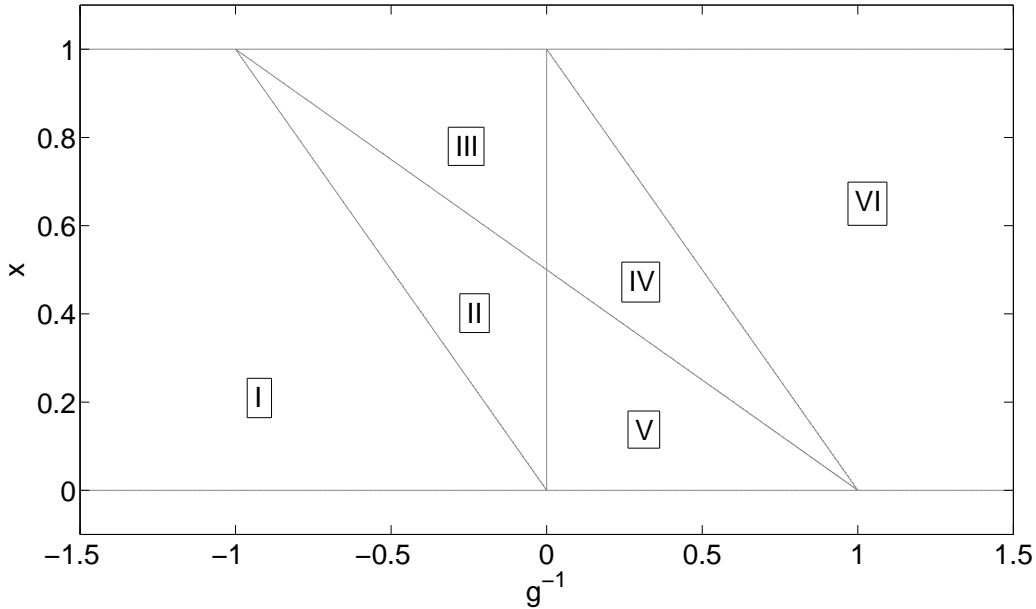}}
\caption{Phase diagram. Regions II and III are dual with respect to the relation (\ref{duality}), as are regions IV and V. Regions III and IV are dual with respect to the relation (\ref{mixed}) as are regions II and V. These dualities combine to give a duality between regions III and V with respect to the relation (\ref{combined}). Our analysis suggests that the mappings from II to V and from III to IV through (\ref{map0}) are injections, mappings from III to II and IV to V through (\ref{map00}) are surjections, such the the composed mappings from III to V through (\ref{map}) are bijections.}
\end{figure}

\section{Mean-field approximation}

Our final point of discussion concerns a mean-field approximation analysis, which is a standard technique applied to the analysis of pairing Hamiltonians in general. It has been previously shown in \cite{dilsz10,rdo10} that the mean-field gap and chemical potential equations and the Bethe Ansatz solution (\ref{bae1}) for the ground state in the continuum limit are equivalent. Here we investigate the extension of that correspondence to include the Bethe Ansatz Eqs. (\ref{bae2}).      

%WORK THROUGH THIS PROPERLY - KEY OBSERVATION IS INDEPENDENCE OF SIGN OF G

%Comparisons between exact results and those obtained by use of a mean-field approximation have been made previously %\cite{x}. Here we recall the relevant details. 
Using a mean-field approach, in particular where products of operators $A$ and $B$ are approximated as 
$$
AB\approx A\langle B \rangle +\langle A \rangle B -\langle A \rangle \langle B \rangle 
$$ 
the Hamiltonian (\ref{ham}) may be approximated by  

\begin{equation}
\mathcal{H}=H_{0}-\frac{1}{2}\hat{\Delta}^*Q-\frac{1}{2}\hat{\Delta}Q^{\dagger}+\frac{\Delta^{2}}{4G}-\mu(N- M), \label{hamiltonianmean}
\end{equation}
where $\hat{\Delta}=2G\left\langle Q \right\rangle$, $\Delta=|\hat{\Delta}|$, $N$ is the particle-pair number operator, $M=\langle N \rangle$ and $\mu$ is a Lagrange multiplier which is introduced since the mean-field approximation does not conserve particle number. 
Setting
\begin{align*}
\mathcal{E}(z_{j})=\sqrt{( z^2_{j}-\mu)^{2}+z_{j}^{2}\Delta^{2}} 
%\label{spectra}
\end{align*}
the ground-state energy is found to be  
\begin{equation}
E_{\rm min}=\frac{1}{2}\sum_{j=1}^L( z^2_{j}-\mu)-\frac{1}{2}\sum_{j=1}^L\mathcal{E}(z_{j})+\frac{\Delta^{2}}{4G}+\mu M \label{energymean}
\end{equation}
associated to the mean-field ground state  
\begin{equation}
\begin{aligned}
\left|\Psi_{\rm min}\right\rangle&=\prod_{j=1}^L(u_{j}I+v_{j}b_{j}^{\dagger})  \left|0\right\rangle \\
&=\prod_{j=1}^L(u_{j}b_j+v_{j}I)  \left|\chi\right\rangle
\label{mfs1}
\end{aligned}
\end{equation}
where
\begin{align*}
|u_{j}|^{2}=\frac{1}{2}\left(1+\frac{z^2_{j}-\mu}{\mathcal{E}(z_{j})}\right),\quad |v_{j}|^{2}=\frac{1}{2}\left(1-\frac{z^2_{j}-\mu}{\mathcal{E}(z_{j})}\right). 
\end{align*} 
Through use of the Hellmann-Feynman theorem we may take partial derivatives of (\ref{hamiltonianmean}) and 
(\ref{energymean}) to generate the following constraint equations:
\begin{align}
\frac{1}{G}&=\sum_{j=1}^L\frac{z_{j}^{2}}{\mathcal{E}(z_{j})} , \label{hf1} \\
L-2M&=\sum_{j=1}^L\frac{z^2_{j}-\mu}{\mathcal{E}(z_{j})} \label{hf2} 
\end{align}
which are known as the gap and chemical potential equations. It is apparent that (\ref{hf1}) cannot admit a solution when $G<0$. However (\ref{bae1}) maps to (\ref{bae2}) with the change $G\rightarrow -G$ and changing the quantum number to count hole-pairs instead of particle-pairs, while the mean-field wavefunction ({\ref{mfs1}) can be equally expressed in terms of particle creation operators acting on the vacuum or particle annihilation operators (i.e. hole creation operators) acting on the completely filled particle state. At first sight it appears there is a paradox.  

If on the other hand we calculate that highest energy state of the approximation (\ref{hamiltonianmean}) we find 
\begin{equation}
E_{\rm max}=\frac{1}{2}\sum_{j=1}^L( z^2_{j}-\mu)+\frac{1}{2}\sum_{j=1}^L\mathcal{E}(z_{j})+\frac{\Delta^{2}}{4G}+\mu M \label{energymean1}
\end{equation}
associated to the mean-field highest-energy state  
\begin{equation}
\begin{aligned}
\left|\Psi_{\rm max}\right\rangle&=\prod_{j=1}^L(v^*_{j}I-u^*_{j}b_j^\dagger)  \left|0\right\rangle. \\
&=\prod_{j=1}^L(v^*_{j}b_j-u^*_{j}I)  \left|\chi\right\rangle.
\label{mfs2}
\end{aligned}
\end{equation}
where $*$ denotes complex conjugation. Through use of the Hellmann-Feynman theorem we may take partial derivatives of (\ref{hamiltonianmean}) and 
(\ref{energymean1}) to generate the following constraint equations:
\begin{align}
-\frac{1}{G}&=\sum_{j=1}^L\frac{z_{j}^{2}}{\mathcal{E}(z_{j})} , \label{hf11} \\
2M-L&=\sum_{j=1}^L\frac{z^2_{j}-\mu}{\mathcal{E}(z_{j})}. \label{hf22} 
\end{align}
Eqs. (\ref{hf11},\ref{hf22}) are equivalent to (\ref{hf1},\ref{hf2}) via the particle-hole transformation and the change $G\rightarrow -G$. This basically asserts that the mean-field approach is justified in calculating the low energy spectrum of the attractive model or the high energy spectrum of the repulsive model. The observation is entirely consistent with (\ref{transform}). Note that in either case we can project the mean-field states (\ref{mfs1},\ref{mfs2}) onto the sector with fixed $M$, which leads to the following unnormalised states
\begin{align}
\left|\Psi_{\rm min}\right\rangle\rightarrow \left( \sum_{j=1}^L \frac{v_j}{u_j}b_j^\dagger \right)^M|0\rangle
&\propto \left( \sum_{j=1}^L \frac{u_j}{v_j}b_j   \right)^{L-M}|\chi\rangle,  \label{proj1}\\
\left|\Psi_{\rm max}\right\rangle \rightarrow \left( \sum_{j=1}^L \frac{u^*_j}{v^*_j}b_j^\dagger \right)^M|0\rangle
&\propto \left( \sum_{j=1}^L \frac{v^*_j}{u^*_j}b_j   \right)^{L-M}|\chi\rangle.
\label{proj2}
\end{align}

It is known \cite{dilsz10} that (\ref{proj1}) is exactly the ground state on the boundary between regions IV and VI (Moore-Read line). Both forms shown in (\ref{proj1}) are obtainable by the Bethe Ansatz solutions (\ref{bae1}) and (\ref{bae2}) respectively. By the same methods it can be shown that (\ref{proj2}) is exactly the highest energy state on the boundary between regions I and II. These boundary lines provide the most ``mean-field-like'' states. In the hole-pair picture they arise from mappings of the form (\ref{map0}) where the domain is the one-dimensional space with basis $\{\\chi\rangle\}$.

What is not apparent is how a mean-field approach may be implemented to observe the structure of states with the form (\ref{map}) in region V, even at the level of an approximation. Conversely there are well-established methods (e.g. see \cite{admor02}) which in principle permit the calculation of the ground state energy of the attractive system from the Bethe Ansatz solution (\ref{bae2}) in the continuum limit. Whether or not this approach simply reproduces the continuum limit of (\ref{hf1},\ref{hf2}), or produces some new insights, presents an interesting open question.

\section{Conclusion}

We presented an alternative form  of the Bethe Ansatz equations, based on the hole-pair picture, in order to re-examine the $p+ip$ model. One of the main results we discovered is an intrinsic asymmetry between the particle-pair and the hole-pair perspectives, in contrast to the $s$-wave paring model on which such a symmetry can be imposed \cite{plm13,pb15}. In particular for the attractive pairing system there are instances of diverging roots of the Bethe Ansatz equations in the hole-pair picture which can be precisely identified and counted. It led us to conjecture that all states in the strong pairing regime have the form of a quasi-condensate with  the same number of zero energy pairs, and infinite energy pairs whose energy sum is finite. Significantly, diverging roots do not occur in the Bethe Ansatz solution of the attractive model in the particle-pair picture. Our findings are summarised in the phase diagram Fig. 2. A notable feature of the phase diagram is that the boundary lines, which were determined by exact calculation without approximation, and all independent of the parameters $z_l$ implicit in the Hamiltonian (\ref{ham}).

\section*{Acknowledgements}
Jon Links and Amir Moghaddam were supported by the Australian Research Council through Discovery Project DP110101414. Jon Links also received support through Discovery Project DP150101294. Ian Marquette was supported by the Australian Research Council through Discovery Early Career Researcher Award DE130101067.

\section*{Appendix A - Direct calculation of the exact solution}

We start with the observation that 
 $$H|\chi\rangle=\sum_{j=1}^L z_j^2|\chi\rangle. $$
To determine exact eigenstates of the Hamiltonian by way of a Bethe Ansatz solution, we follow the approach of \cite{bil12}. Define generic states of the form  
\begin{align*}
 |\psi\rangle=\prod_{k=1}^M B(y_k)|\chi\rangle, \qquad
 |\psi_j\rangle=\prod_{k\neq j}^M B(y_k)|\chi\rangle, \qquad 
 |\psi_{jl}\rangle=\prod_{k\neq j,l}^M B(y_k)|\chi\rangle 
\end{align*} 
where $B(y)$ is given by (\ref{by}).
Noting the commutation relations
\begin{align*}
\left[H_0,\,B(y)\right]
&=-\sum_{j=1}^L\frac{z^3_j}{y-z_j^2} b_j  \\
&= Q-yB(y) \\
\left[Q^\dagger,\,B(y)\right]&=\sum_{j=1}^L\frac{z_j^2}{y-z_j^2}(2N_j-I)  
\end{align*}
it is then found that
\begin{align*}
(H-\sum_{j=1}^Lz_j^2I)|\Psi\rangle&=((1-G)H_0+(G-1)\sum_{j=1}^Lz_j^2I)|\Psi\rangle -G Q Q^\dagger|\Psi\rangle \\
&=  (1-G)\sum_{j=1}^P\left(Q|\Psi_j\rangle-y_j|\Psi\rangle\right)  -GQ  \sum_{j=1}^P\sum_{p=1}^L\left(B(y_1)...\left(\frac{z_l^2}{y-z_l^2}(2N_p-I)\right)...B(y_P)  \right)|\chi\rangle \\
&=  (1-G)\sum_{j=1}^P \left(Q|\Psi_j\rangle-y_j|\Psi\rangle\right)  +G Q\sum_{p=1}^L\sum_{j=1}^P\frac{z_p^2}{y_j-z_p^2} |\Psi_j\rangle \\
& \qquad \qquad  +2 GQ\sum_{j=1}^P\sum_{r>j}^P\sum_{p=1}^L \frac{z_p^3}{(y_j-z_p^2)(y_r-z_p^2)}b_p^\dagger|\Psi_{rj}\rangle \\
&= (1-G)\sum_{j=1}^P\left(Q|\Psi_j\rangle-y_j|\Psi\rangle\right) -G Q \sum_{j=1}^P\sum_{l=1}^L\frac{z_l^2}{y_j-z_l^2}|\Psi_j\rangle \\
&\qquad \qquad +G Q\sum_{j=1}^P\sum^P_{r\neq j}\sum_{l=1}^L \left(\frac{ y_r{z_l^2}}{(y_j-y_r)(y_r-z_p^2)}+\frac{y_j{z_l^2}}{(y_r-y_j)(y_j-z_p^2)}\right)b_l|\Psi_{rj}\rangle \\
&= (1-G)\sum_{j=1}^P\left(Q|\Psi_j\rangle-y_j|\Psi\rangle\right) -G Q\sum_{l=1}^L\sum_{j=1}^P\frac{z_l^2}{y_j-z_l^2}|\Psi_j\rangle\\
&\qquad \qquad + GQ\sum_{j=1}^P\sum^P_{r\neq j} \left(\frac{y_r}{y_j-y_r}|\Psi_j\rangle+\frac{y_j}{y_r-y_j}|\Psi_r\rangle\right)   \\
&= (1-G)\sum_{j=1}^P\left(Q|\Psi_j\rangle-y_j|\Psi\rangle\right) -G Q\sum_{j=1}^P\sum_{l=1}^L\frac{z_l^2}{y_j-z_l^2}|\Psi_j\rangle  +2 GQ\sum_{j=1}^P\sum_{r\neq j}^P \frac{y_r}{y_j-y_r}|\Psi_j\rangle    
\end{align*}
The terms proportional to $|\Psi_j\rangle$ cancel provided
\begin{align*}
G-1+G \sum_{l=1}^L\frac{z_l^2}{y_j-z_l^2}&=2G \sum^P_{r\neq j} \frac{y_r}{y_j-y_r}, \qquad
k=1,...,P 
\end{align*}
which can be equivalently written as (\ref{bae2}). 
%\begin{align}
%\frac{-G^{-1}+2P-L-1}{y_k} + \sum_{l=1}^L\frac{1}{y_k-z_l^2}& =\sum^P_{j\neq k}\frac{2}{y_k-y_j},
%\qquad k=1,..., P.  
%\label{baenew}
%\end{align}
For each solution of those coupled equations, $|\psi\rangle$ is an eigenstate of the Hamiltonian with energy eigenvalue given by 
$$ E=\sum_{l=1}^Lz_l^2+(G-1)\sum_{k=1}^P y_k. $$

\section*{Appendix B - Proof of an inequality} 
Here we show that when $M+M'=L-G^{-1}$ with $G^{-1}>0$, and $L\geq M' \geq M$, that 
\begin{align}
d(M')\geq d(M)
\label{theorem}
\end{align}
where $d(M)$ is given by (\ref{dim}). Noting that 
\begin{align*}
L\geq M' \geq L-M'-G^{-1},
\end{align*}
this is equivalent, in view of (\ref{dim}), to establishing that for $A-C\geq B \geq C$, 
\begin{align*}
\frac{C!(A-C)!}{B!(A-B)!}\geq 1.
\end{align*}

When $X>Y$ we have from the definition of the factorial that  
\begin{align*}
\frac{X!}{Y!}&\leq X^{X-Y}, \\
\frac{X!}{Y!}&\geq(Y+1)^{X-Y}.
\end{align*}
Then
\begin{align*}
\frac{(A-C)!}{B!}&\geq (B+1)^{A-B-C}, \\
\frac{C!}{(A-B)!}&\geq\frac{1}{(A-B)^{A-B-C}}.
\end{align*}
This establishes that 
\begin{align*}
\frac{C!(A-C)!}{B!(A-B)!}\geq 1
\end{align*}
whenever $2B+1\geq A$. Alternatively,
\begin{align*}
\frac{(A-C)!}{(A-B)!}&\geq (A-B+1)^{B-C}, \\
\frac{C!}{B!}&\geq\frac{1}{B^{B-C}}.
\end{align*} 
This establishes that 
\begin{align*}
\frac{C!(A-C)!}{B!(A-B)!}\geq 1
\end{align*}
whenever $A+1\geq 2B$. It then follows that (\ref{theorem}) is true.

\end{document}